\documentclass[prb,aps,twocolumn,groupaddress,longbibliography]{revtex4-1}
\usepackage{latexsym}
\usepackage{lineno}
\usepackage{hyperref}
\usepackage{url}
\usepackage{graphicx}
\usepackage{verbatim}
\usepackage{multirow}
\usepackage{amsmath}
\newcommand{\beq}{\begin{eqnarray}}
\newcommand{\eeq}{\end{eqnarray}}
\usepackage{mathrsfs}
\usepackage{float}
\usepackage[usenames, dvipsnames]{color}
\usepackage{mathtools}
\usepackage{slashed}
\usepackage{physics}	
\usepackage{graphicx}   
\usepackage{epstopdf}
\usepackage{subfigure}  
\usepackage{hyperref}   
\usepackage{bbold}
\usepackage{wasysym}
\usepackage{amssymb}
\usepackage{feynmp}
\usepackage[symbol]{footmisc}

\usepackage[latin1]{inputenc}
\usepackage{tikz}
\usetikzlibrary{decorations.pathmorphing}
\usetikzlibrary{shapes.misc}
\tikzset{cross/.style={cross out, draw=black, minimum size=8*(#1-\pgflinewidth), inner sep=0pt, outer sep=0pt},
cross/.default={1pt}}
\usetikzlibrary{patterns,math}
\begin{document}

\title{Anharmonic theory of superconductivity in the high-pressure materials}

\author{Chandan Setty}
\thanks{csetty@ufl.edu}
\affiliation{Department of Physics, University of Florida, Gainesville, Florida, USA}
\author{Matteo Baggioli}
\thanks{matteo.baggioli@uam.es}
\affiliation{Instituto de Fisica Teorica UAM/CSIC, c/Nicolas Cabrera 13-15,
Universidad Autonoma de Madrid, Cantoblanco, 28049 Madrid, Spain.}
\author{Alessio Zaccone}
\thanks{alessio.zaccone@unimi.it}
\affiliation{Department of Physics ``A. Pontremoli", University of Milan, via Celoria 16, 20133 Milan, Italy.}
\affiliation{Cavendish Laboratory, University of Cambridge, JJ Thomson Avenue, CB30HE Cambridge, U.K.}

\begin{abstract}
Electron-phonon superconductors at high pressures have displayed the highest values of critical superconducting temperature $T_c$ on record, now rapidly approaching room temperature. 
Despite the importance of high-$P$ superconductivity in the quest for room-temperature superconductors, a mechanistic understanding of the effect of pressure and its complex interplay with phonon anharmonicity and superconductivity is missing, as numerical simulations can only bring system-specific details clouding out key players controlling the physics. Here we develop a minimal model of electron-phonon superconductivity under an applied pressure which takes into account the anharmonic decoherence of the optical phonons. We find that $T_c$ behaves non-monotonically as a function of the ratio $\Gamma/\omega_0$, where $\Gamma$ is the optical phonon damping and $\omega_0$ the optical phonon energy at zero pressure and momentum. Optimal pairing occurs for a critical ratio $\Gamma/\omega_0$ when the phonons are on the verge of decoherence (``diffuson-like'' limit). Our framework gives insights into recent experimental observations of $T_c$ as a function of pressure in the complex BCS material TlInTe$_2$. 
\end{abstract}
\maketitle

\section{Introduction} 
When a crystal lattice is subjected to a (hydrostatic) pressure deformation, its phonon frequencies change in response to the change of volume, in a way which is controlled by the materials's Gr{\"u}neisen parameter, hence by the anharmonicity of the vibration modes.
However, the effects of these changes in the phonon frequencies, and of the related anharmonicity, on the superconducting properties of a material have largely remained poorly understood. 
Filling this knowledge gap is an urgent problem in order to develop an understanding of superconductivity in materials under pressure, which include the highest-$T_c$ values recorded so far in the high-pressure hydride materials~\cite{Eremets2015,Pickett,Eremets2019,Eremets_review}. 

On one hand, a large number of experimental works have shown how the superconducting critical temperature $T_c$ changes as a function of pressure $P$ for a variety of materials. 
For elemental superconductors, a commonly observed trend in experiments is a decrease of $T_c$ with increasing $P$, which has been theoretically predicted upon analyzing the behaviour of the Eliashberg electron-phonon coupling function $\alpha^{2} g(\omega)$ as a function of $P$, see Refs.\cite{Carbotte1,Carbotte2,Zavaritskii,Hodder,Ginzburg}. 
A similar behaviour is seen in many technologically important materials, such as Nb$_{3}$Sn, see Ref.~\cite{Jaccard}.
An increase of $P$ typically shifts the $\alpha^{2} g(\omega)$ distribution to higher frequencies, thus driving the system into an unfavorable regime as per the Bergmann-Rainer criterion~\cite{Bergmann1973} for the electron-phonon coupling. 
Notable exceptions to the above standard rule for elemental superconductors is represented by $\alpha$-uranium~\cite{Gardner,Ginzburg}, while another puzzling material such as bismuth is known to have a very low $T_c$ at ambient pressure (on the order of the mK)~\cite{Prakash,Behnia} and a decent $T_c$ ($7-8$K) at higher pressures~\cite{Wittig}. In both these systems, new effects play a role. In $\alpha$-uranium the phonon density of states is very rich of soft vibrational modes~\cite{Manley}, traditionally attributed to anharmonicity in crystals~\cite{baggioliPRL}, although their origin in $\alpha$-uranium is still under debate~\cite{Manley}. In bismuth, instead, the Debye energy is very close to the Fermi energy, thus leading to an almost vanishing attraction for the Cooper pairs~\cite{Behnia}. At high pressure, a more close-packed structure becomes favourable, which changes the underlying phonon physics leading to more favourable conditions for pairing.

On the other hand, exploring the effect of pressure on more complex non-elemental materials has led to a zoology of trends of $T_c$ as a function of $P$, see e.g. Ref.~\cite{Chu}. In such materials, exceptions to canonical $T_c$ behavior with pressure according to phonon frequency shifts and the Bergmann-Rainer criterion lay abound. In almost all cases, understanding and isolating the key ingredients that affect $T_c$ as a function of pressure at the level of model Hamiltonians is a futile exercise given their incredible microscopic complexity. The cuprates form a case in point where even the pairing mechanism is highly debated and the quasiparticle picture is ill-defined. Nevertheless, in phonon mediated superconductors, numerical simulations have provided invaluable quantitative insights into the phonon dispersion relations, and into the structural stability of superconducting compounds, including many new materials. 

More specifically, numerical calculations allow one to estimate the anharmonicity of the various phonon modes involved, by comparing fully anharmonic calculations with harmonic calculations \cite{Mauri2015,Eremets_review}. As a matter of fact, early theoretical approaches~\cite{Carbotte1,Carbotte2,Zavaritskii,Hodder,Ginzburg} ignored phonon anharmonicity while other approaches~\cite{Ganguly,Maksimov,Mahan1996}, including more recent works on the high-$T_c$ hydrides~\cite{Mauri2015,Bergara2010, Mauri2013, Mauri2014, Mauri2016, Arita2016, Errea2016, Bergara2016,szcesniak2015high,kostrzewa2020lah,errea2020quantum,camargomartnez2020higher} focus mainly on the phonon energy renormalizations neglecting anharmonic damping/decoherence. 
The latter is a key ingredient that is needed to properly describe the effect of pressure on phonon-mediated Cooper pairing, as will be shown in our work. Thus a mechanistic picture of high pressure effects on the superconducting state which exploits the synergy between anharmonic phonon decoherence and phonon energy renormalization is missing in most materials including elemental superconductors.  

In this paper, we develop a minimal version of such a theory by working with a gap equation in the weak coupling BCS limit. Crucially, to mediate Cooper pairing, we implement optical phonon propagators which contain the effect of an external applied pressure and the resulting anharmonic decoherence via the optical phonon damping~\cite{Klemens1966,Setty2020}. The analytical theory is able to provide predictions that allow one to disentangle the complex interplay between pressure-induced changes of optical phonon energy and anharmonic decoherence, and their effects on the $T_c$. The results of our theory are presented in the specific context of a recent high pressure study on the superconductor TlInTe$_2$.  Different physical regimes are predicted, which include (i) monotonic decrease of $T_c$ with $P$ as observed in many systems; (ii)  non-monotonic trend with a minimum, in conjunction with optical phonon softening, which qualitatively explains recent experiments in  TlInTe$_2$ from Ref.\cite{yesudhas2020origin}; (iii) non-monotonic trend with a maximum in a regime of incoherent phonons where the quasiparticle picture breaks down. 

We emphasize that our goal is to approach this question from a phenomenological viewpoint with key inputs from experiments. We do not wish to provide accurate predictions of $T_c$ as a function of pressure -- an endeavor elusive to even state-of-art numerical methods. Rather, we take the perspective that the coordination between decoherence and frequency renormalization induced by phonon anharmonicity \textit{can} play a role dominant enough to provide a reasonable qualitative understanding of experimental data. From such a proof-of-principle demonstration, our expectation is that this synergy between energy scales must necessarily constitute a key ingredient of any serious future numerical first principle study that aims to understand superconducting properties of materials such as TlInTe$_2$. 

\section{Experimental standpoint}
Recent high pressure Raman spectroscopy, X-ray diffraction and transport measurements in TlInTe$_2$ along with first-principles band structure calculations uncovered a change in Fermi surface topology due to a Lifshitz transition between 6.5-9 GPa, leading to the formation of enlarged electron pockets at the Fermi level. This feature is preceded by a superconducting transition at 5.7 GPa with $T_c\simeq 4K$.  With increasing pressure, the $T_c$ decreases steadily and rises again with a minimum located around 10 GPa. Concurrent to this V shaped $T_c$ ``anomaly'', there is a further softening of the $A_g$ phonon mode. It is natural to attribute such a V-shaped $T_c$ behavior to changes in electronic density of states or the softening of the $A_g$ phonon mode as  was concluded by the authors of Ref.~\cite{yesudhas2020origin}. After all, both these quantities play key roles in controlling superconducting properties especially in phonon mediated superconductors. 

But on closer examination, these arguments are  debatable at best.  First, the theoretically calculated electronic density of states (DOS) from the electron pocket becomes larger in the regime between 6GPa and 9GPa due to the Lifshitz transition. However, this is exactly the regime where $T_c$ decreases, thus eliminating the electronic DOS as the key driver for the observed trend in $T_c$. Second, the softening of the $A_g$ phonon mode occurs around $P^*\sim12.5$ GPa greater than the pressure where $T_c$ is minimum.  But any argument justifying a decrease in $T_c$ with increasing phonon frequency  implicitly invokes the Bergmann-Rainer criterion. While this criterion works well for $P<P^*$, the same argument fails when $P>P^*$ since a softening phonon mode would imply a second dip in $T_c$ approximately symmetric with respect to $P^*$. This however seems to contradict experimental observations.     

Having ruled out a dominant role of (purely) phonon frequency shifts or electronic DOS in explaining observed experimental behavior, we turn to the possibility that the phonon linewidth could be a key player in determining superconducting properties in TlInTe$_2$. Raman data as a function of pressure indicates that anharmonicity in this material is strong enough to significantly increase the phonon linewidth ($\Gamma$) with respect to the peak frequency $\omega'$, the latter controlled by the key parameter  $\omega_0$, but weak enough so that the phonons remain coherent ($\Gamma < \omega_0$). As we will see below, this is precisely the regime where $T_c$ correlates with the ratio $\Gamma/\omega_0$.  The delicate balance in hierarchy of scales and the possibility of disentangling other effects such as electronic DOS and phonon frequency shifts, makes TlInTe$_2$ an ideal playground to test the hypothesis presented in this paper.  
We note that the experimental situation in TlInTe$_2$ is evolving. For example, it is still not clear whether the normal state yielding the superconductor exhibits all conventional Fermi liquid properties. To date, there are no attempts to determine the pairing symmetry of the superconducting state either. Even the nature and full symmetry characterization of phonons responsible for pairing, the strength of individual electron-phonon couplings etc,  is undetermined. \textcolor{black}{Furthermore, there exist uncertainties between experiment and theory in the measurement of the bulk modulus and its derivative,  which could in principle change the finer details of the relationship between pressure and frequency. So far, we know of no improved equation of state that  accounts for all the experimental measurable quantities precisely,  so we use the best candidate available in literature, i.e. the well-known Birch-Murnaghan equation of state for deriving the relationship between pressure and frequency (used also in the experimental study of Ref.~\cite{yesudhas2020origin}).  Hence we keep our formalism simple but general enough to accommodate these uncertainties until future experiments paint a more complete picture of the material's phenomenology.}

\begin{figure}
    \centering
    \includegraphics[width=1.0 \linewidth]{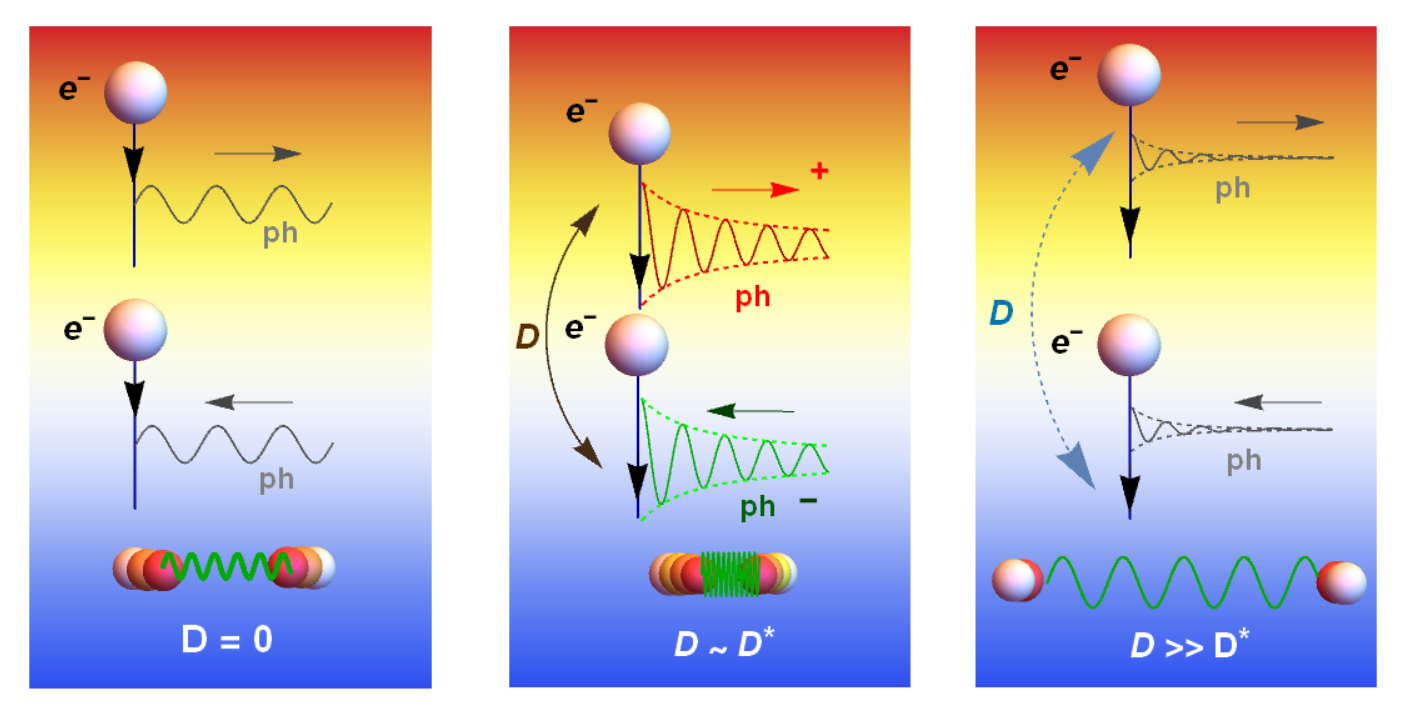}
    \caption{Mechanism of $T_c$ enhancement through anharmonicity with two phonon modes. \textbf{(Left panel)} In the absence of anharmonic decoherence ($D=0$), the Stokes (S-) and anti-Stokes (aS-)  processes are insensitive to their phase and are thus indistinguishable. This scenario leads to ordinary Cooper pairing. \textbf{(Middle panel)} Weak anharmonic decoherence ($D \sim D^*$) sensitizes the phase of the S- and aS- processes and enables them to act coherently and enhance the effective coupling of electrons and phonons leading to strong Cooper pairs. \textbf{(Right panel)} For very strong anharmonicity ($D\gg D^*$), the S- and aS- processes are only weakly sensitive to their phases making them effectively indistinguishable while acting to reduce the effective coupling of electrons and phonons leading to weak pairing.}
    \label{fig:sketch}
\end{figure}

\section{Theoretical framework} 
\subsection{Optical phonon energy under pressure}
We start by analyzing the effect of external pressure on the optical phonons of a crystal lattice. The main effect of pressure is to induce a negative volume change of the material. The change of volume, in turn, is related to a change of phonon frequency, through the Gr{\"u}neisen parameter, $\gamma = - d \ln \omega' / \ d \ln V$, via~\cite{Kunc}:
\begin{equation}
\frac{\omega'(V)}{\omega'_{P=0}}=\left(\frac{V}{V_0}\right)^{-\gamma}\,,
\label{Gruneisen}
\end{equation}
where $\omega'_{P=0}$ refers to optical phonon energy at zero ambient pressure.
The above relations apply to individual phonon modes with frequency $\omega'$. 

The volume change is related to the change of pressure as described by the Birch-Murnaghan equation of state~\cite{Birch}, which is derived based on nonlinear elasticity theory, and provides an expression for $P(V)$. Upon replacing $V$ with $\omega'$ in \eqref{Gruneisen}, one obtains the following relation between the optical phonon frequency $\omega'$ and the applied pressure~\cite{Birch}:
\begin{equation}
    P(X)\,=\,\frac{3}{2}\,B_0\,\left(X^{7}-X^{5}\right)\,\left[1+\eta\,(1-X^{2})\right]\label{kunc}\,,
\end{equation}
with $X \equiv (\omega'/\omega'_{P=0})^{1/3\gamma}$. Upon inverting the above Eq.\eqref{kunc} to obtain $\omega'$ as a function of $P$, it is clear that $\omega'$ is a monotonically increasing function of $P$ in the regime of interest here, with the increase being modulated by anharmonicity through $\gamma$. Also, $B_0$ is the bulk modulus, while $\eta=(3/4)(B_{0}'-4)$ with $B_{0}'=d B_{0}/dP$.

In the above relations, the frequency $\omega'$ refers to the real part of the phonon dispersion relation (which already contains the renormalization shift due to anharmonicity~\cite{Baowen}), whereas the imaginary part of the dispersion relation is related to the phonon damping coefficient $\Gamma$ (the inverse of the phonon lifetime), as follows (e.g. Eqs. (23)-(27)in Ref.~\cite{Baowen})
\begin{eqnarray}
    \omega^2&\,=\,&\omega_0^2\,-\,i\,\omega\,\Gamma\,+\,\mathcal{O}(q^2),\\ 
    \omega' &\equiv& \mathrm{Re}(\omega)\,=\,\frac{1}{2}\,\sqrt{4\,\omega_0^2\,-\,\Gamma^2}\,+\mathcal{O}(q^2),\quad \\
    \frac{\Gamma}{2}&\,\equiv\,&\mathrm{Im}(\omega)\,+\mathcal{O}(q^2).
\label{optical}
\end{eqnarray}
Quantitative numerical calculation of $\Gamma$ can be done using the Self-Consistent Phonon (SCP) methodology~\cite{Baowen,Tadano1}, for specific systems~\cite{Tadano2}, but this is not the goal of our paper, which is rather focused on generic qualitative trends in terms of the effect of $\Gamma$ on the pairing and on $T_c$.
Hence, $\omega'$ denotes the renormalized phonon energy measured e.g. in Raman scattering (i.e. the Raman shift), while $\Gamma$ represents the linewidth of the Raman peak. Let us emphasize that these expressions are at leading order in the momentum $q$ and higher order corrections $\mathcal{O}(q^2)$ are neglected at this stage.

We now introduce a key dimensionless parameter for the subsequent analysis
\begin{equation}
D\equiv \Gamma /\omega_0\,,
\end{equation}
which quantifies the degree of coherence of the phonon. Low values of $D$ signify high coherence of the phonons, which can thus be treated as approximately independent quasiparticles, whereas, at the opposite end of the spectrum, very large $D$ values correspond to incoherent vibrational excitations in the diffusive regime (``diffusons'' in the language introduced by Allen, Feldman and co-workers~\cite{Allen_diffusons}).
The schematic picture that will emerge from the subsequent theoretical analysis is anticipated in Fig.\ref{fig:sketch}.

In the following section, we introduce the theoretical framework for the Cooper pairing and we will start by considering how the superconducting critical temperature $T_c$ varies as a function of $D$. 

\subsection{Gap equation with anharmonic phonon damping}
 For a generic Fermionic Matsubara frequency $\omega_n$ and momentum $\textbf k$, we denote the gap function as $\Delta(i\omega_m,\textbf k)$. We assume throughout a quadratic dispersion relation for the electronic band. With a constant coupling $g$, the gap equation can be derived from the Eliashberg equations in the one-loop ( weak coupling) approximation, and takes the form~\cite{Marsiglio,Kleinert}
\begin{eqnarray}\nonumber
\Delta(i\omega_n, \textbf{k}) &=& \frac{g^2}{\beta V} \sum_{\textbf{q}, \omega_m} \frac{\Delta(i\omega_m, \textbf{k}+ \textbf{q}) \Pi(\textbf{q}, i\omega_n - i \omega_m)}{\omega_m^2 + \xi_{\textbf{k} + \textbf{q}}^2 + \Delta(i\omega_m, \textbf{k}+ \textbf{q})^2}\,,\\
&&
\label{Sum-GapEqn}
\end{eqnarray}
where $\beta$ is the inverse temperature and $V$ is the volume. In Matsubara frequency space, we choose the pairing mediator to be a damped optical phonon given by the bosonic propagator~\cite{Ziman}
\begin{equation}
\Pi(\textbf{q}, i \Omega_n) = \frac{1}{\Omega(q)^2 + \Omega_n^2 + \Gamma(q)\Omega_n},
\label{propagator}
\end{equation}
where $\Omega_n$ is the bosonic Matsubara frequency, $\Omega(q) = \omega_0 + \hat{a} \,q^2$ is the phonon dispersion, and the damping factor, $\Gamma(q) \equiv D\omega_0$, is a constant independent of momentum for  high-frequency optical phonons~\cite{Klemens1966}. In accordance with the Klemens formula~\cite{Klemens1966}, one can also include an additional prefactor, $1+\frac{2}{e^{\omega_0/2T}-1}$, in the damping term $\Gamma(q)$ to account for a temperature dependent phonon linewidth. We find that this has a negligible effect on the results discussed below. The factor $D$ controls the strength of the damping term and may change with pressure.  The leading order contribution to the square of the  dispersion is $\Omega(q)^2 \simeq \omega_0^2 + v q^2$ where $v = 2 \omega_0 \hat{a}$. This is the first momentum correction which was neglected in Eq.\eqref{optical}. Assuming an isotropic, frequency-independent gap $\Delta(i\omega_n, \textbf k)\equiv\Delta$, we can set the external frequency and momentum to zero without any loss of generality (see supplementary note~\cite{Note} with regards to the $\omega_n=0$ simplification). 
Converting the resulting summation into an energy integral (and assuming a quadratic dispersion relation for the fermions), the gap equation becomes \\
\begin{eqnarray} \nonumber
1 &=& \sum_{\omega_m} \int_{-\mu}^{\infty}\frac{\lambda T d\xi}{\left[ v \xi + M^2 + \omega_m^2 - D \omega_m\omega_0\right] \left[\omega_m^2 + \xi^2 + \Delta^2\right]}\,,\\
&&
\label{Integral-GapEqn}
\end{eqnarray}
where $M^2 = \mu v + \omega_0^2$. Here we have defined the effective coupling constant $\lambda = N(0)g^2$, $N(0)$ is the density of states at the Fermi level, and $\mu$ is the chemical potential. We can now utilize the energy integral identity $\int_{-\infty}^{\infty} \frac{d \xi}{(z \xi + s) (\xi^2 + r^2)} = \frac{\pi s}{r(s^2 + z^2 r^2)}$ in the limit of large chemical potential to yield the gap equation
\begin{eqnarray}\nonumber
    1 &=& \sum_{\omega_m}\frac{\lambda \pi T \left(M^2 + \omega_m^2 - D \omega_m \omega_0\right)}{\sqrt{\omega_m^2 + \Delta^2}\left[(M^2+\omega_m^2 - D\omega_m\omega_0)^2 + (\omega_m^2 + \Delta^2)v^2\right]}\,.\\
    &&
    \label{MSum}
\end{eqnarray}
We can now perform the final Matsubara sum using methods described in Ref.~\cite{Varlamov2005} after seeking a condition for $T_c$ by setting $\Delta=0$. Defining $p = \omega_0 D + i v$ and $Q_{\pm} = \frac{1}{2}\left(p \pm \sqrt{p^2 - 4 M^2}\right)$ leads to an equation for $T_c$ that can be numerically solved given by
\begin{widetext}
\begin{eqnarray}
-M'^2 &=& \psi\left(\frac{1}{2}\right) + \frac{1}{4}\Bigg[\left\{ \frac{p'-Q_+'}{Q_+' - Q_-'}\psi\left(\frac{1}{2}-\frac{Q_+'}{2\pi T_c'}\right)+\frac{-p'+Q_-'}{Q_+' - Q_-'}\psi\left(\frac{1}{2}-\frac{Q_-'}{2\pi T_c'}\right) + c.c\right\} + \left\{ D \rightarrow -D \right\}\Bigg],
\label{gap}
\end{eqnarray}
\end{widetext}
where $\psi\left(x\right)$ is the digamma function, the primed quantities are dimensionless and are defined as $Q_{\pm}' \equiv \frac{Q_{\pm}}{\sqrt{\lambda}}$, $T_c' = \frac{T_c}{\sqrt{\lambda}}$, and so on.

\section{Results}
\subsection{Schematic $T_c$ dependence on optical phonon energy and anharmonicity}
\begin{figure}[t]
    \centering
    \includegraphics[width=1.0\linewidth]{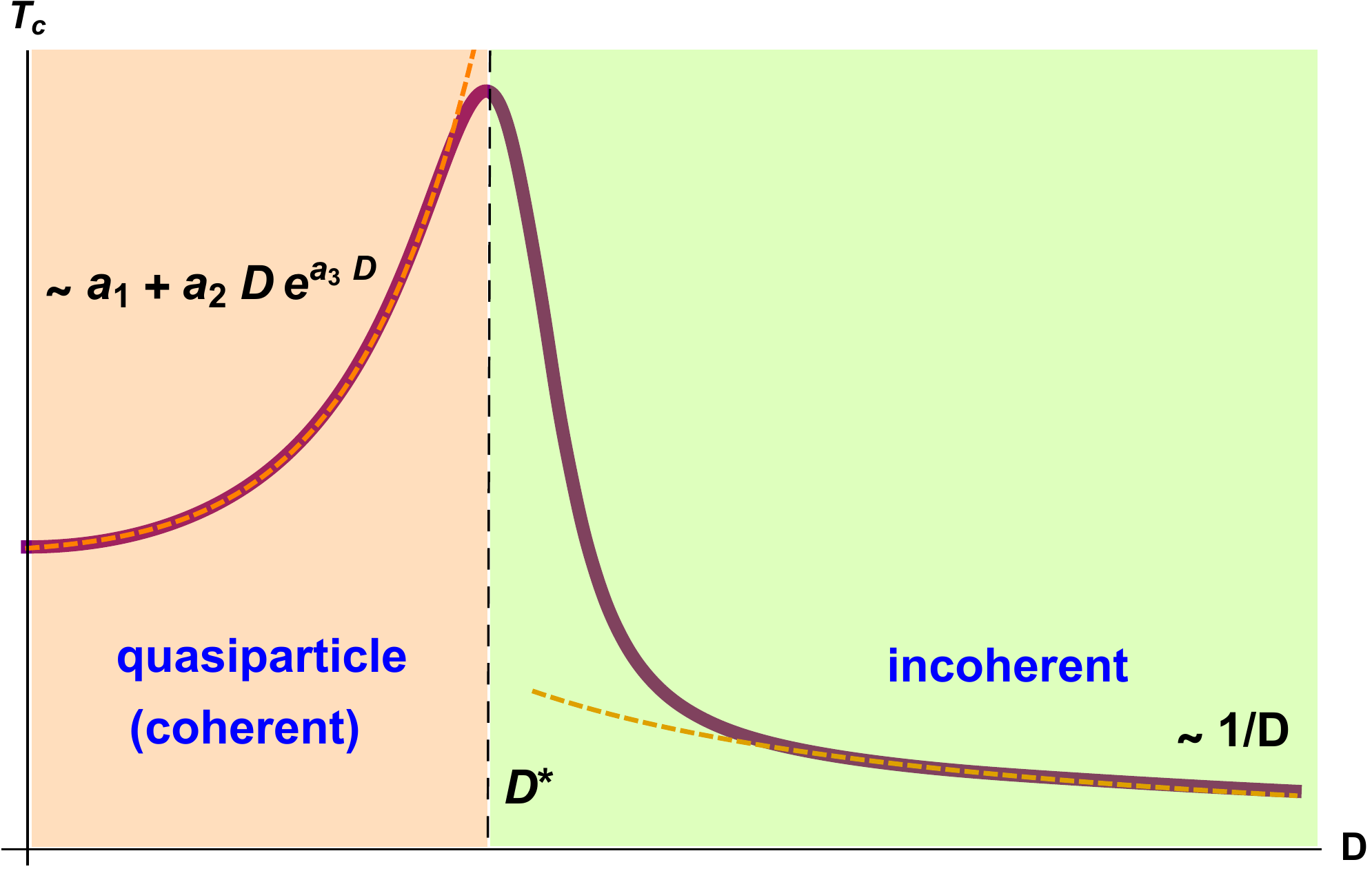}
    \caption{An illustration of the two regimes present in our model: the ``coherent'' regime where the critical temperature grows with $D \equiv \Gamma/\omega_0$ and the ``incoherent'' regime, where the functional dependence is inverted. In the ``coherent'' regime, the optical phonons behave like independent quasiparticles with frequencies renormalized by anharmonicity, whereas in the incoherent regime the quasiparticle coherence breaks down due to the large anharmonic damping.}
    \label{fig0}
\end{figure}
Upon numerically solving Eq.\eqref{gap} for a constant damping coefficient $\Gamma$, we can study the evolution of $T_c$ as a function of the dimensionless parameter $D\equiv \Gamma/\omega_0$. The trend is shown in Fig.\ref{fig0}.
At low $D$ values, $T_c$ increases with $D$, then goes through a maximum after which it then decays sharply upon further increasing $D$. \color{black} The maximum appears around $D^* \sim \mathcal{O}(1)$ with its exact value determined by the microscopic parameter $M^2$. This corresponds exactly to the scale at which the real and the imaginary part of the phonon dispersion relation become comparable ($\Gamma \sim \omega_0$) and the phonons turn into quasi-localized ``\textit{diffuson-like}'' excitations~\cite{Allen_diffusons}. In this sense, this is analogous to the Ioffe-Regel crossover scale \cite{ioffe1960non}. \color{black}

The mechanistic picture shown in Fig.\ref{fig:sketch} can be used to understand the non-monotonic dependence of $T_c$ upon the anharmonic decoherence parameter $D$. To begin, we note that in the absence of $D$, the gap equation in Eq.~\ref{Integral-GapEqn} has even terms only in the Matsubara frequency transfer $\omega_m$. Hence both constructive Stokes (S-) and destructive anti-Stokes (aS-) processes, which emit and absorb energy respectively, contribute to the gap equation equivalently. However, when $D$ is non-zero, Eq.~\ref{Integral-GapEqn} is sensitive to the sign of the energy transfer, thereby distinguishing the two processes. From this property, it is clear that the energy integral and Matsubara summations in Eqs.~\ref{Integral-GapEqn} and~\ref{MSum} lead to terms that are proportional to $D$ in the numerator of the gap equation. Provided $D\lesssim D^*$, this effectively increases the electron-phonon coupling  $\lambda$ and hence the Cooper pair binding energy. For values of $D$ much larger than $D^*$, the phonons are extremely damped and S- and aS- processes again contribute approximately equally to the gap equation, thus reducing the effective electron-phonon coupling. \\

At low values of damping $\Gamma$ (low-$D$ regime) and $\Gamma/\omega_{0} \ll1$, the real part of the dispersion relation dominates over the imaginary part, and the phonons behave like coherent quasiparticles with well-defined momentum $k$. In the opposite regime of large anharmonic damping $\Gamma/\omega_{0} \gg 1$ (hence large $D$), we have that $\mathrm{Im}\omega>\mathrm{Re}\omega$, hence the phonons lose their coherence and the quasiparticle approximation breaks down. These two regimes correspond to two different Cooper pairing regimes. One regime we call the ``coherent'' regime (because here phonons behave like coherent quasiparticles), where $T_c$ correlates positively with anharmonic damping (hence where damping enhances $T_c$).  The second regime we call  ``inchoerent'' and here, instead, $T_c$ decreases with further increasing the anharmonic damping. 
Notice that, in the coherent regime, $T_c$ increases (decreases) as the optical phonon energy $\omega_0$ decreases (increases), whereas the opposite trends apply in the incoherent regime. 
This implies that the effect of pressure can either promote or depress superconductivity depending on the underlying physics of the optical phonons in a given lattice. In the Appendix, we provide additional plots for the variation of $T_c$ with other physical parameters $M'$ and $v'$ appearing in Eq.~\ref{gap}. For low and high $D$, $T_c$ is barely affected by $M'$ and $v'$ whereas the peak value of $T_c$ is suppressed at critical $D^*$ with increasing values of both these parameters. On the other hand, the peak $D^*$ itself increases with $M'$ while it is barely affected by $v'$.   \\

The theoretical prediction in Fig.\ref{fig0} can be fitted with the following simple functions
\begin{align}\nonumber
     T_c(D)&\,\sim\,&a_1+a_2 \,D\, e^{a_3\,D}\,\quad \textit{for}\quad D < D^{*}\,\,( \textit{coherent}),\\
      T_c(D)&\,\sim\,&D^{-1}\, \quad \textit{for}\quad D > D^{*}\quad (\textit{incoherent})\,,
\end{align}
 with $a_n>0$.\\

We will show below that these two regimes lead to radically different scenarios in terms of the dependence of $T_c$ on the external pressure $P$. This conceptual schematization will be shown in the next sections to hold a number of consequences for a deeper mechanistic understanding of the effect of pressure on superconductivity in complex materials.

In Fig.\ref{fig0} we assumed that the pairing is mediated by high-frequency optical phonons near the Debye frequency $\omega_{D}$ for which the Klemens model gives a simplified (constant) anharmonic damping coefficient $\Gamma = D \omega_0$.
In the more general case, the Klemens damping is given by $\Gamma= \alpha \omega_0^5$, where $\alpha$ is a prefactor which depends on the microscopic physics which governs the decay of the optical phonon into two acoustic phonons. Notably, $\alpha \sim \gamma^{2}$, where $\gamma$ is the lattice Gr\"uneisen parameter introduced above. The latter is a function of the interatomic potential~\cite{Krivtsov2011}, hence of the electronic orbital/bonding physics, and can be easily computed, for a given phonon mode in a given material, from first principles~\cite{Bongiorno}. 
 
\begin{figure}[t]
    \centering
    \includegraphics[width=0.9\linewidth]{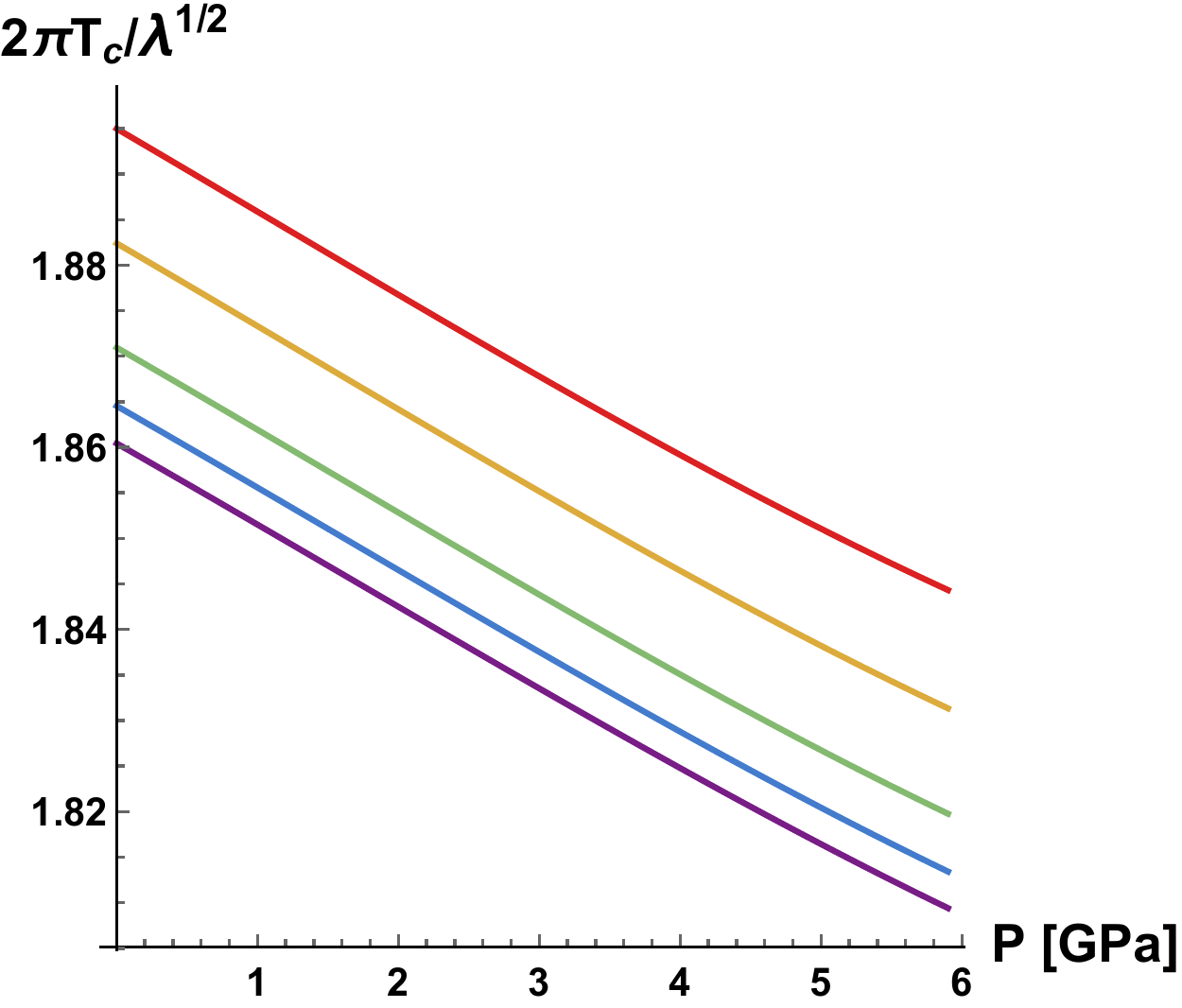}
    \caption{The normalized critical temperature with $D= \alpha \,\omega_0^4$ and $\omega_0$ given by the formula Eq.\eqref{kunc}. $\alpha$ decreases from purple to red, $\alpha=\{0.5,0.48,0.45,0.4,0.35\}\,\times\,10^{-7}$. The $y$-axis is dimensionless in our units. $\omega_0(P)$ is taken from the experimental fit shown in Fig.4(a).}
    \label{finalb}
\end{figure}

Using this more general Klemens formula for a generic optical phonon that mediates the pairing, we obtain the trends shown in Fig.\ref{finalb}. A linear decreasing trend of $T_c$ as a function of $P$ is predicted by our theory for the incoherent-phonon (strongly anharmonic) regime. A linearly decaying trend of $T_c$ with $P$ has been recently observed in the strongly anharmonic AlH$_{3}$ high-pressure hydride~\cite{Errea2021} as well as in the SC-I phase of CeH$_{10}$ in Ref.~\cite{Oganov}.
In more standard systems, a linear decay of $T_c$ with increasing $P$ has been reported for in the literature for simple (e.g. elemental) superconductors~\cite{Ginzburg,Palmy,Chu}.

\subsection{Theoretical analysis of superconductivity in TlInTe$_2$ at high pressure}
In this section, we explore the potential of the above framework to rationalize recent experimental data where highly non-trivial (e.g. non-monotonic) dependencies of $T_c$ upon $P$ have been observed, and for which a theoretical explanation is lacking. 
We study the paradigmatic case of TlInTe$_2$, for which accurate experimental data are available for the phonon mode $A_g$ involved in the Cooper pairing. Data are available in terms of the optical phonon energy and of the anharmonic damping, as measured by Raman scattering, and also for $T_c$, as a function of pressure~\cite{yesudhas2020origin}.

\begin{figure}[t]
    \includegraphics[width=0.9\linewidth]{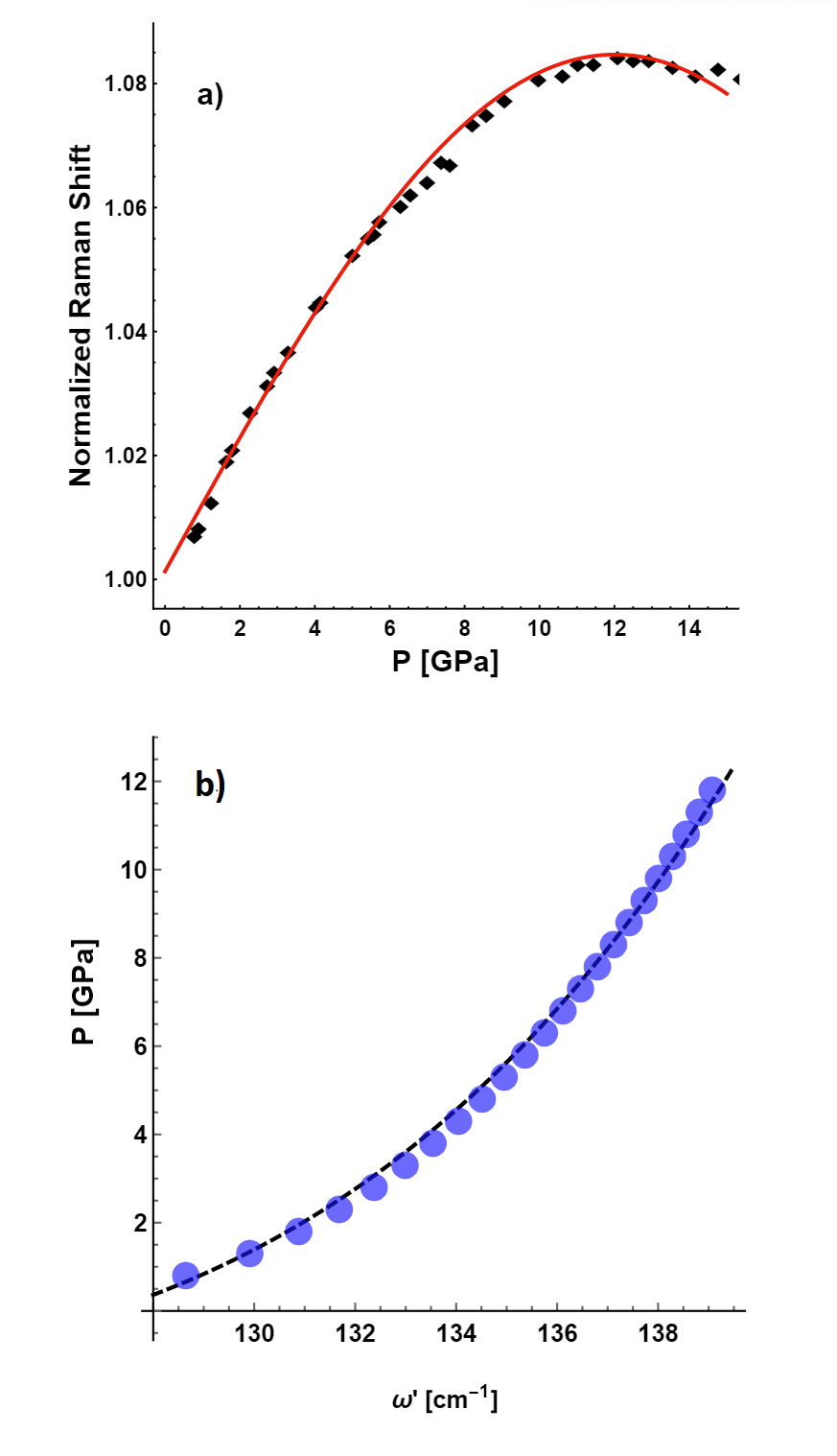}
    \caption{\textbf{(a) } The normalized Raman shift (proportional to $\omega'$) of the $A_g$ phonon mode in  TlInTe$_2$ as a function of pressure and its fit with an empirical function. The value at zero pressure is $\approx 128$ cm$^{-1}$. Data taken from \cite{yesudhas2020origin}. \textbf{(b) } Comparison between the best empirical fit of \cite{yesudhas2020origin} (shown in the panel (a)) and the Eq. \ref{kunc} in terms of $\omega_0\approx \omega'$, ($\Gamma \ll \omega_0$). The parameters are set to the values shown in Eq.\eqref{param}.}
    \label{fig3}
\end{figure}

We start by fitting the experimental data for the frequency of the Raman-active $A_g$ optical phonon (renormalized by anharmonicity) $\omega'$ as a function of pressure, displayed in Fig.\ref{fig3} (a). By using Eq.\ref{kunc} for the fitting, we get:
\begin{eqnarray} \nonumber
    &B_0\,=\,15\,\textrm{GPa}\,,\qquad \gamma\,=\,0.3\,,  \\ 
    &\eta\,=\,-2.475, \qquad \omega'_{P=0}\approx\omega_{0,P=0}\,=\,127\,\textrm{cm}^{-1}. \label{param}
\end{eqnarray}
where we fixed $\eta = -2.475$ as found experimentally from the P-V relation in Ref.~\cite{yesudhas2020origin}.
The value that we found for $\gamma$ is close to value found for the $A_g$ mode in  this material~\cite{yesudhas2020origin}, $\gamma \sim 0.23$, and larger values (up to $0.8$) were also reported in the literature~\cite{Ves}. Also the value of the bulk modulus that we found from our fitting (15GPa) is quite close to the experimental value (19GPa) reported in Ref.~\cite{yesudhas2020origin}.

The fitting is shown in Fig.\ref{fig3}(b), where the frequency values refer to $\omega'$. The latter has been obtained by using Eq.\ref{optical} in combination with Eq.\ref{kunc}. 
The optical mode energy increases upon increasing $P$ in a conventional way~\cite{Kunc} up to $P= 8$GPa, after which phonon softening, linked to the increase of anharmonic damping $\Gamma$ is observed upon further increasing $P$, as shown in Fig. \ref{fig3}(a). 

\begin{figure}[t]
    \centering
     \includegraphics[height=0.7\linewidth]{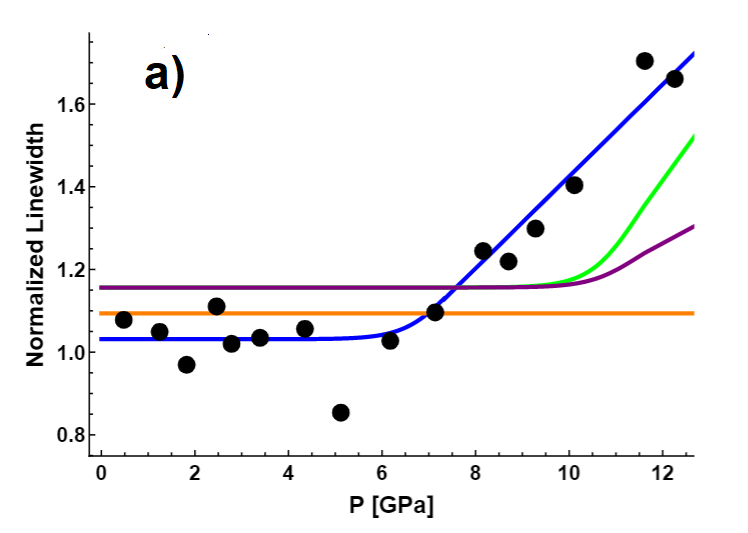}
     
     \vspace{0.2cm}
     
      \includegraphics[height=0.7\linewidth]{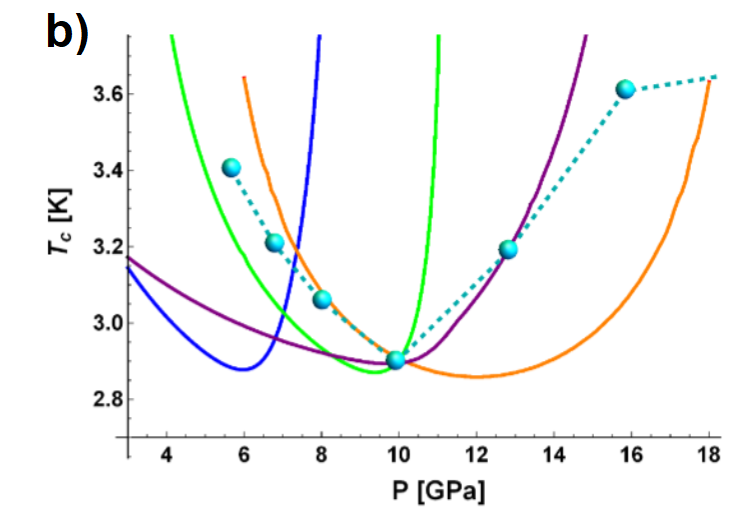}
     
    \caption{\textbf{(a) }The normalized linewidth of the $A_g$ phonon mode in TlInTe$_2$ and three different sets of fits. The zero pressure value is taken as $3.2$ cm$^{-1}$. Data taken from \cite{yesudhas2020origin}. \textbf{(b) }The corresponding theoretical calculations for the critical temperature (solid lines) are compared with the experimental data (symbols). The colors of the theoretical curves for $T_c$ match the respective models for the linewidth in the panel (a). The parameters used in the model correspond to $\hat{a}=1$, $ \alpha = (6,5.4,5.4,4.2)\times 10^{-8}$eV$^{-4}$, $\mu' =\frac{\mu}{\sqrt{\lambda}}= (37.3, 32.5, 35.6, 24)$ for orange, blue, green and purple curves. We also choose $\lambda=1$, $\omega_0 \sim 15$meV $= \frac{1}{2} v$.  } 
    \label{fig4}
\end{figure}

The increase of anharmonicity with pressure is clearly evidenced by the behaviour of the Raman peak linewidth $\Gamma$, as shown in Fig.\ref{fig4}(a). Notice that the percentile growth of the linewidth under pressure is much larger than that of the normalized Raman shift. In this sense, the material is characterized by giant anharmonicities and the damping effects are fundamental. 
Here, in the same panel, different empirical trends are shown, alongside the experimental data which manifest a significant scatter. 
In general, $\Gamma \ll \omega_0$ for this system, such that this case belongs to the ``coherent'' regime discussed in the previous section and in Fig.\ref{fig0}. Indeed, we checked that $\omega'$ and $\omega_0$ differ by only about $0.01\%$ at all $P$ values.
These different trends for $\Gamma$ have been implemented, alongside the fitted optical phonon energy $\omega'$ from Fig.\ref{fig3}, into our theoretical gap-equation framework for the prediction of $T_c$ presented in the previous section of this paper.
The resulting theoretical $T_c$ trends are shown in Fig.\ref{fig4} in comparison with the experimental $T_c$ data from Ref.~\cite{yesudhas2020origin}, as a function of the applied pressure.

All the $\Gamma$ trends in Fig. \ref{fig4} (a) clearly lead to the same qualitative dependence of $T_c$ on $P$, with a minimum. The physics behind this trend is explained by our theoretical framework: at low $P$ the $T_c$ decreases because of the increase in $P$, which induces an increase of the optical phonon frequency $\omega'$ or $\omega_0$. The subsequent phonon softening leads to the minimum and to an inversion of the trend: upon further increasing the pressure the $T_c$ starts to rise.  This is due to the fact that lower $\omega_0$ values lead to a Stokes/anti-Stokes constructive interference (in the presence of anharmonic damping), which enhances the Cooper pairing~\cite{Setty2019, SBZ2020}. This behaviour, with a minimum in $T_c$ is independent of the particular $\Gamma$ trend with $P$, and in fact occurs even for $\Gamma$ constant with $P$. 

The role of the $\Gamma$ trend with $P$ is to control the position of the minimum as a function of pressure. Also, importantly, the presence of a rise in $\Gamma$ leads to a stronger rise after the minimum, which confirms that in the ``coherent'' regime the $T_c$ can be strongly enhanced by the anharmonic damping, as discussed in the context of Fig.\ref{fig0}. This finding has deep implications for high-$T_c$ hydrogen-based materials, where the anharmonic damping of the optical phonons can be significant and may be tuned by the material design.
Also, phonon softening could also be enhanced by the electron-phonon interaction itself as discussed in Refs.~\cite{Marsiglio2020,Schmalian}.

\section{Conclusion}
We presented a theory of the pressure effect on Cooper pairing in superconductors where the pairing is mediated by generic bosonic excitations.
Our theory is based on solving the gap equation with a bosonic propagator that is damped due to anharmonic decoherence. A specific calculation is presented for optical phonons which takes into account: (i) the anharmonicity of the phonon via the Klemens' damping, (ii) the effect of pressure on the phonon frequency. The theory identifies two fundamental regimes as a function of the dimensionless ratio $D$ between anharmonic phonon damping and phonon frequency. At low values of this ratio, $T_c$ is strongly enhanced by anharmonicity and, at the same time, decreases with increasing pressure. At large values of the $D$ ratio (after a maximum), where the phonons are no longer well-defined quasiparticles, $T_c$ instead correlates positively with pressure and is lowered by anharmonicity (see Fig.\ref{fig0}). Optimal pairing occurs for a critical ratio $D^*$ when the phonons are on the verge of decoherence (``diffuson'' limit).

A linearly decreasing correlation between $T_c$ and P ~\cite{Chu} is predicted to occur in the regime of strongly anharmonic phonons.
Furthermore, the theory provides a qualitative description of recent experimental data on TlInTe$_2$ for which phonon frequencies, anharmonic phonon damping and $T_c$ were all measured experimentally.
It predicts that $T_c$ initially decreases with $P$ as a consequence of the optical phonon energy increasing with $P$, but then  goes through a minimum, as the optical phonon starts to soften and to become more anharmonic, after which it rises with $P$. The predicted behaviour is well supported by the experimental data.

This theoretical picture provides a mechanistic rationale for the pressure effect on superconductivity in TlInTe$_2$, by physically describing different regimes of negative/positive pressure effect on $T_c$. By clarifying the deep interplay between anharmonicity of the bosonic glue and pressure effects on the pairing mechanism, the theory provides new guidelines for material design, which may prove useful for discovering and/or engineering new materials with enhanced $T_c$. In future work, the presented framework could be combined with models of strain-dependent critical properties in technologically important materials such as Nb$_{3}$Sn, where anharmonic phonon generation has been recently shown to play a key role~\cite{Valentinis}.\\

\textit{Acknowledgements} M.B. acknowledges the support of the Spanish MINECOas ``Centro de Excelencia Severo Ochoa'' Programme under grant SEV-2012-0249. C.S. is supported by the U.S. DOE grant number DE-FG02-05ER46236. A.Z. acknowledges financial support from US Army Research Laboratory and US Army Research Office through contract nr. W911NF-19-2-0055. 

\onecolumngrid
\section*{Appendix}
In Fig. \ref{fig3}, the functional form for the
empirical fits of the phonon linewidth is given by the following expression:
\begin{equation}
     f(x)^{\text{fit}}\,=\,  0.00064035\, x^5-0.00398124 \,x^3+1.39603\, x+128.166
\end{equation}
This fitting function works well up to $P \approx 15$ GPa. Interestingly it seems to capture also the turning of the data point between $15<P<25$ but it definitely fails in capturing the last two points of the dataset. 

In Fig. \ref{fig4}, the Raman linewidth is normalized with respect to the experimental zero pressure value, taken as $3.2 \,cm^{-1}$. 
The curves shown in the panel (a) of Fig.\ref{fig4} are given by:
\begin{align}
    & \text{orange:}\,\,\,\,\,\Tilde{\Gamma}(P)\,=\,1.09375\,,\\
    &\text{blue:}\,\,\,\,\,\Tilde{\Gamma}(P)\,=\,0.3125 \left(
\begin{array}{cc}
  & 
\begin{array}{cc}
0.09523 \,(0.4422 \,P+0.0597) (1-\tanh (7.326\, -P))+3.3 & P\leq 7.326 \\
 0.356 \,P+1 & 7.326<P \\
\end{array}
 \\
\end{array}
\right)\\
&\text{green:}\,\,\,\,\,\Tilde{\Gamma}(P)\,=\,0.3125 \left(
\begin{array}{cc}
 & 
\begin{array}{cc}
0.09523 (0.4422 \,P+0.0597) (1-\tanh (11.326\, -P))+3.7 & P\leq 11.326 \\
 0.5\, P-1.47 & 11.326<P \\
\end{array}
 \\
\end{array}
\right)\\
&\text{purple:}\,\,\,\,\,\Tilde{\Gamma}(P)\,=\,0.3125 \left(
\begin{array}{cc}
 & 
\begin{array}{cc}
0.09523 (0.1822 \,P+0.0597) (1-\tanh (11.326\, -P))+3.7 & P\leq 11.56 \\
 0.1952\, P+1.7 & 6<P \\
\end{array}
 \\
\end{array}
\right)
\end{align}
Additionally, the parameters for the panel (b) are as follow:
\begin{align}
    &\text{orange:}\,\,\,\,\,\alpha\,=\,6\,\times\,10^{-8}\,\,,\,\,\mu'\,=\,37.3\,,\\
    &\text{blue:}\,\,\,\,\,\alpha\,=\,5.4\,\times\,10^{-8}\,\,,\,\,\mu'\,=\,32.5\,,\\
    &\text{green:}\,\,\,\,\,\alpha\,=\,5.4\,\times\,10^{-8}\,\,,\,\,\mu'\,=\,35.6\,,\\
    &\text{purple:}\,\,\,\,\,\alpha\,=\,4.2\,\times\,10^{-8}\,\,,\,\,\mu'\,=\,24\,.
\end{align}
where we set $\hat{a}=1$ (the optical phonon stiffness) and $\mu' = \frac{\mu}{\sqrt{\lambda}}$. This corresponds to $\mu \sim 10eV$ for a BCS superconductor with $\omega_0\sim 15$ meV $= \frac{1}{2}v$.  We have checked that the dependence of $\omega'$ on $P$ is predominantly controlled by the $P$-dependence of $\omega_0$, whereas the contribution of the $P$-dependence of $\Gamma$ in the square root appearing in the expression for $\omega'$ is much smaller (which is also consistent with our conclusions that the experimental data of TlInTe$_2$ are largely in the weak-anharmonicity coherent-phonon regime). 
\\ \newline
\begin{figure}
    \centering
    \includegraphics[width=0.45 \linewidth]{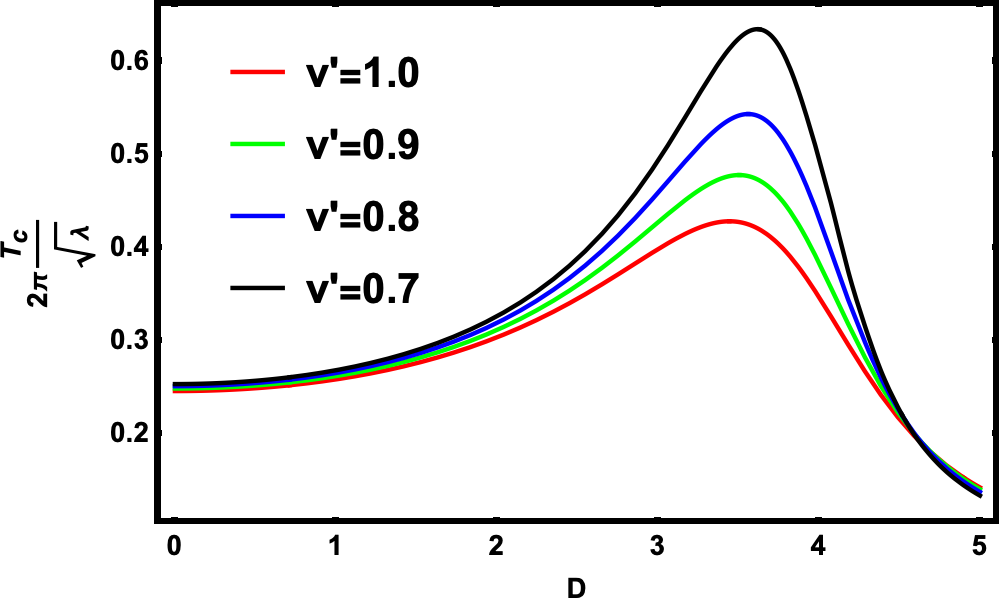}\qquad
    \includegraphics[width=0.45 \linewidth]{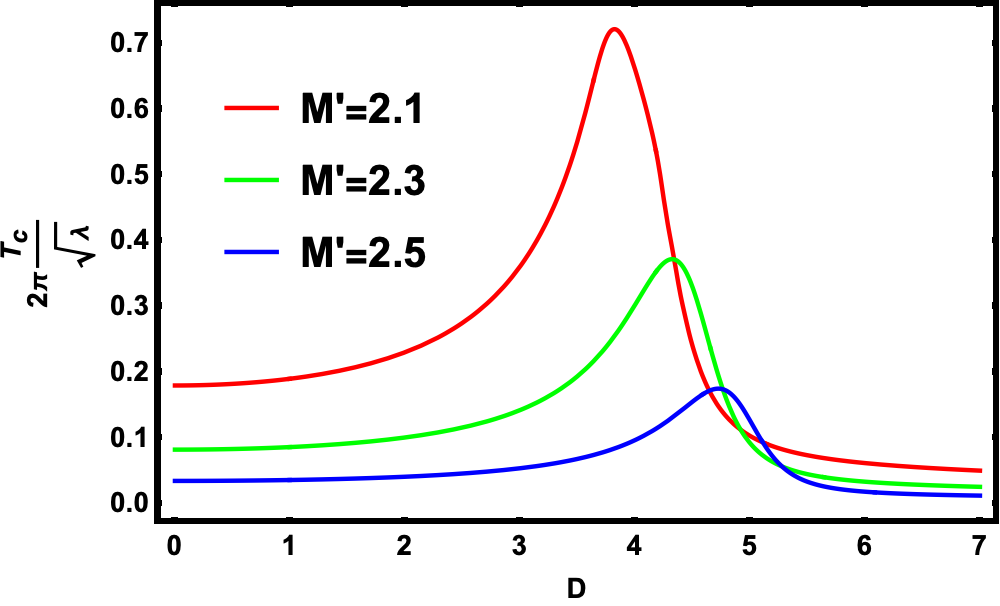}
   \caption{Plot of the dimensionless critical temperature $T_c$ as a function of $D$ for various $v'$ (left) and $M'$ (right). We have chosen $M'=2$ (left panel) and $v'=0.5$. The Klemens' factor has a negligible effect in both cases.  }
    \label{TcVsParameters}
\end{figure}
In Fig.~\ref{TcVsParameters}, we also plot the dimensionless $T_c$ as a function of $D$ for various $v'$ and $M'$. As stated in the main text, increasing both $v'$ and $M'$ reduces the peak $T_c$. While $v'$ barely has an effect on the critical $D^*$, increasing $M'$ increases $D^*$.

\twocolumngrid

\bibliography{anharmonicity2}

\begin{thebibliography}{59}
\bibitem [{Not()}]{Note}%
  \BibitemOpen
  \href@noop {} {\bibinfo  {journal} {In general circumstances, the
  simplification of setting $\omega_n=0$ without accounting for the additional
  ``1/2" would yield incorrect results and qualitatively different predictions.
  However, we have checked that this is not the case here. The peak $T_c$
  enhancement is overestimated in the simplified solution but the qualitative
  behavior of the non-monotonicity is unaffected. This minimal cost for
  simplification yields a more tractable formula (there would be two
  unsymmetrized terms without the simplification) that can be fit to
  experiments more readily}\ }\BibitemShut {NoStop}%
\end{thebibliography}

\end{document}